\documentclass[epj]{svjour}
\usepackage{graphics}
\begin{document}

\def\mco{\multicolumn}

\newcommand{\beq}{\begin{equation}}
\newcommand{\eeq}{\end{equation}}
\newcommand{\lb}{\label}
\newcommand{\beqa}{\begin{eqnarray}}
\newcommand{\eeqa}{\end{eqnarray}}

\newcommand{\apol}{\alpha}
\newcommand{\ldB}{\lambda_{dB}}

\newcommand {\Lc} {L_{crit}}
\newcommand {\lc} {\ell_{crit}}
\newcommand {\leff} {\ell_{eff}}
\newcommand {\lw} {\ell_{w}}
\newcommand {\rc} {r_{cl}}
\newcommand {\rw} {r_w}
\newcommand {\rg} {r_g}
\newcommand {\bc} {b_c}
\newcommand {\Ef} {\vec{E}}
\newcommand {\rsq} {1/\rho^2}
\newcommand {\kat} {k_{atom}}

\newcommand {\PRL} {Phys.Rev.Lett.~}
\newcommand {\PR} {Phys.Rev.~}
\newcommand {\ea} {{\em et al.}~}

\newcommand{\infig}[2]{\begin{center}\mbox{\input epsf \epsfxsize#2
\epsfbox{#1}}\end{center}}

\title{Quantum Wires and Quantum Dots for Neutral Atoms}
\author{J\"{o}rg Schmiedmayer\inst{1}
}                     
%
%
\institute{Institut f\"{u}r Experimentalphysik, Universit\"{a}t Innsbruck, A-6020 Innsbruck,  AUSTRIA}
\date{Received: October 1997 / Revised version: date}
%
\abstract{
By placing changeable nanofabricated structures (wires, dots, etc.)
on an atom mirror one can design guiding and trapping potentials
for atoms. These potentials are similar to the electrostatic
potentials which trap and guide electrons in semiconductor quantum
devices like quantum wires and quantum dots.  This technique will
allow the fabrication of nanoscale atom optical devices.
} 
\maketitle
\section{Introduction}
\label{intro}

Cooling and trapping techniques developed in recent years allow for
the preparation of very cold and dense samples of neutral atoms
(for a recent review see \cite{SpIssue}).  These atoms are so slow
that their typical deBroglie wavelength, $\ldB$, can be on the
order of 100~nm or larger which is in the range of the size of
mesoscopic structures built using modern nanofabrication
technology.  Choosing the suitable interactions between the neutral
atoms and nanofabricated structures it should be possible to build
mesoscopic quantum devices with atoms guided and/or trapped in
designed potentials.

In this paper we will first discuss the basic principles of surface
mounted mesoscopic atom optical devices and then describe how to
design {\em quantum wires} and {\em quantum dots} for neutral
atoms.  This will be followed by a short discussion of how to load
atoms in these mesoscopic guides and traps.  Finally we will point
to some of the possible applications of such devices.

\section{Surface mounted atom optics}
\lb{s:SurfAtOpt}

Several different interactions between a neutral atom and
nanofabricated structures such as a thin wire or a sharp tip (dot)
can lead to trapping and guiding.  The principal idea is the
following: An attractive interaction binds the atom to the wire
(dot) and a repulsive interaction, close to the wire (dot) surface,
prevents the atom from interacting with the surface.  This
repulsive interaction is very important because atoms hitting the
surface will either be absorbed or scattered inelastically.  In
most cases this surface will be many orders of magnitude hotter
(typically 300K) compared to the kinetic energy of the cold atoms
($E_{kin} < \mu eV$ corresponding to $T<mK$ temperatures).
Therefore the atoms are, for all practical purposes, lost if they
come in contact with the surface.

In this paper we explore atom optic \cite{AtomOptics} devices which
can be built when combining the strong short range repulsive
interaction of an atom mirror (evanescent wave mirrors
\cite{EvWaveMirror,Cook,Balykin,Esslinger,Seifert,EvMirrProblem} or
magnetic mirrors \cite{MagMirror,Sidorov,Roach}) and the attractive
interaction of a neutral atom in the electric field of a charged
wire (dot). By placing charged nanostructures on (or below) an atom
mirror, one can design guiding and trapping potentials which are
similar to the electric potentials used to trap and guide electrons
in semiconductor quantum devices \cite{QuElDevice} like quantum
wires and quantum dots.

The reflection of atoms at an atom mirror is based on a short range
(exponentially decaying) repulsive potential (decay length $1/\kappa $):
\beq \lb{e:at_mirror}
        U_m(x)=U_0 \exp(-\kappa x)
\eeq
where $x$ is the distance from the mirror surface.  $U_m$ can be
created either by a blue-detuned evanescent wave created by total
internal reflection (evanescent wave mirror
\cite{EvWaveMirror,Cook,Balykin,Esslinger,Seifert,EvMirrProblem})
or the magnetic fields of an alternating magnetic pattern (magnetic
mirror \cite{MagMirror,Sidorov,Roach}).

The strong position dependent {\em attractive} potential required
to build guides and traps can be created by the strong
(inhomogeneous) electric field $\Ef$.  The potential energy for a
neutral atom with electric polarizability $\alpha$ (For simplicity
we will assume $\alpha$ to be a scalar) in the electric field is
then given by
\beq \lb{e:el_pol}
        U_{pol}=-2 \pi \epsilon_0\alpha |E(r)|^2 .
\eeq
For the ground state of an neutral atom $U_{pol}$ is always attractive.

Since the repulsive potential is short range, on the order of the
wavelength of light for the evanescent wave mirror, the attractive
Van der Waals interaction between the atom and the surface is
important \cite{VdW} and has to be taken into account in the
calculations. It leads to a modification of the trapping potential,
drawing the minimum closer to the surface of the atom mirror.

\section{Neutral Atom Quantum Wires}
\lb{s:QuantWire}

\begin{figure}
        \begin{center}\mbox{\input epsf \epsfxsize\columnwidth
\epsfbox{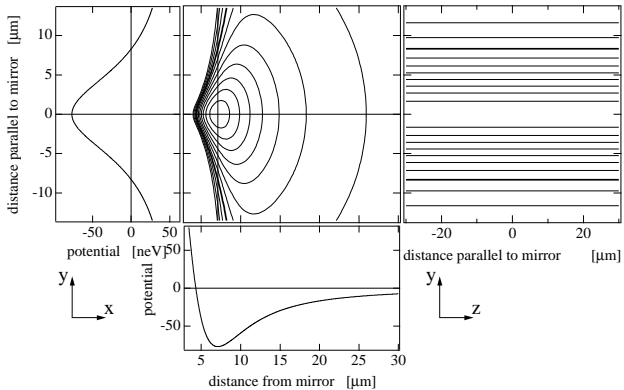}}\end{center}
        \caption{Typical potential for neutral atom quantum wire.  The attractive
        potential ($\rsq$) is created by the interaction of the induced dipole
        moment in the electric field of the charged wire mounted directly on the surface of an atomic
        mirror.  The action of the atomic mirror (evanescent wave or magnetic
        mirror) prevents the atom from reaching the surface and creates a quantum
        channel close to the surface.  The central contour graph shows the potential
        defining the quantum channel.  The two adjacent plots give the potential in
        a direction orthogonal to the charged wire and orthogonal / parallel to the
        mirror surface.  Distances are given from the
        location of the  charged wire and the surface of the atom mirror.}
        \protect{\label{f:qw_potential}}
\end{figure}

A simple mesoscopic one dimensional trapping potential for neutral
atoms can be built by mounting a charged wire on an atomic mirror:

The interaction between a neutral atom and the electric field of a
thin wire with line charge $q$ along the $z$-direction is given by
(in cylindrical coordinates: \\ $\rho=\sqrt{x^2+y^2}, \; \phi , \;
z$ ):
\begin{equation}
        U_{pol}(\rho) =  - \frac{1}{2 \pi \epsilon_{0}}
        \frac{\alpha \, q^2}{2 \rho^2}.
        \label{e:vpol_wire}
\end{equation}
The potential $U_{pol}(\rho)$ is always attractive and diverges
like $\rho^{-2}$ as $\rho \to 0$.  The motion can not be stabilized
by angular momentum and the interaction potential
(Eq.~\ref{e:vpol_wire}) does not lead to stable orbits
\cite{UpolWire}.  Atoms either get absorbed on the wire surface or
they escape towards infinity.  Mounting such a charged wire at the
surface of the atom mirror allows the combination of the attractive
$\rsq$ potential (Eq.~\ref{e:vpol_wire}) with the repulsive
potential of the atom mirror $U_m(x)$ (Eq.~\ref{e:at_mirror}).
\begin{equation}\lb{e:qw_pot}
        U_{guid}(\vec{x})=U_m(x)+U_{pol}(\rho)
\end{equation}
This creates a potential tube for the atoms as shown in
Fig.~\ref{f:qw_potential} which can be viewed as a waveguide for
neutral atoms, the atom optical analog to a gradient index optical
fiber. The interaction potential being equivalent to a refractive
index.

Far above the atom mirror the guiding potential $U_{guid}(\vec{x})$
(Eq.~\ref{e:qw_pot}) is dominated by $U_{pol}(\rho)$  which behaves
like $1/\rho^2$.  Accordingly there is an infinite number of bound
states and the eigenenergies for the high-lying states follow an
exponential law.  However one can adjust the parameters so that
there are only one or a few deeply bound states.  In this regime a
matter wave fiber with quasi single-mode propagation properties can
be built.

\begin{figure}
        \begin{center}\mbox{\input epsf \epsfxsize\columnwidth
\epsfbox{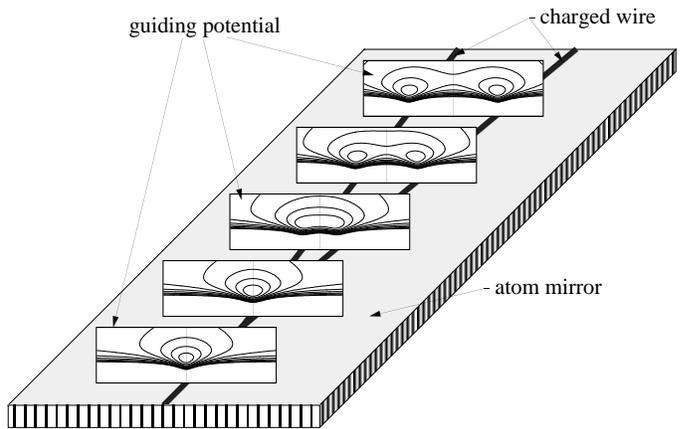}}\end{center}
        \caption{Schematics of a quantum wire beam splitter.  The charged wire
        mounted on an atomic mirror is split in Y shape.  The resulting guiding
        potential is shown in selected planes along the quantum wire beam
        splitter.}
        \protect{\label{f:qwire_BS}}
\end{figure}

Mounting the charged wire on an atomic mirror to create such a
waveguide allows one to use the well-developed nanofabrication
techniques to lithographically write the wires, in any geometrical
structure, i.e.  straight, bent or in form of a Y etc. One could
create beam splitters (see Fig.~\ref{f:qwire_BS}), interferometers
or even complex networks for guided neutral atoms.  Furthermore
additional electrodes located close to the wire can be used to
modify the guiding potential at demand. One can easily imagine
designing switches, gates, modulators etc. for guided atoms.

The above mentioned techniques of manipulating the guiding (trapping)
potentials for neutral atoms is similar to techniques used in quantum
electronics to create quantum wells and quantum wires \cite{QuElDevice},
therefore the name {\em neutral atom quantum wires}.

Typical parameters for such neutral atom guides are given in
table~\ref{t:QuantWire}.  The quantum wires based on the evanescent
wave mirror have much tighter confinement and are closer to the
mirror surface. This is mainly due to the much shorter decay length
of the evanescent wave potential ($\sim 0.1\mu$m) as compared to
the magnetic interaction ($\sim 1.5\mu$m) in ref.~\cite{Roach}.
Secondly the spontaneous scattering rate in the evanescent wave
based quantum wires is quite high (typically a few kHz for a simple
mirror using internal reflection).  This rate can be reduced by
increasing the detuning of the evanescent wave, but then higher
laser power is needed for the same mirror potential height.  This
may be achieved by confining the mirror, and therefore the light,
to only a small region around the wire by guiding the laser light
in a planar, transversely-confined waveguide which may even enhance
the evanescent wave. With such special nanofabricated light guides
producing the repulsive evanescent wave a reduction of the
scattering rate by 2-4 orders of magnitude should be feasible.

The guides formed by the magnetic mirrors are much further above
the surface and can be made much deeper due to the stronger
repulsive mirror potential.  In general the magnetic mirror based
quantum wires seem to be more promising for coherent waveguides,
especially since choosing special magnetic materials with higher
magnetization and with smaller magnetic structures
\cite{r:HindsPriv} for the magnetic mirror will allow for deeper
and more confined quantum wires.  One can also estimate the
lifetime of atoms in these quantum wires due to tunneling to the
mirror surface, and found it to be much longer than 1000~s in any
of the discussed cases.

Using highly charged wires the lower-lying eigenstates in the
neutral atom quantum wire can be localized in transverse direction
to much smaller than the wavelength of light used to manipulate the
atoms (i.e. 670 nm for Li atoms). Atoms will be trapped in the
Lamb-Dicke regime. With better atomic mirrors it is even
conceivable to separate the bound states by more than the linewidth
of the optical transition allowing sideband cooling of neutral
atoms similar to ions in ion traps \cite{iontraps}.

\section{Neutral Atom Quantum Dots}
\lb{s:QuantDot}

In analogy to the above discussed quantum wires, microscopic traps
(the analog to quantum dots) can be created by mounting a charged
tips (point) at or close beneath the atom mirror surface.

\begin{figure}
        \begin{center}\mbox{\input epsf \epsfxsize\columnwidth
\epsfbox{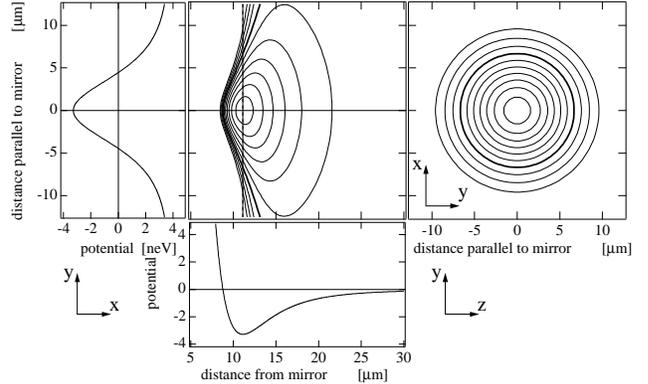}}\end{center}
        \caption{Typical potential for neutral atom quantum dot.  The attractive
        potential ($1/r^{-4}$) is created by the interaction of the induced dipole
        moment in the electric field of a point charge mounted directly on the surface of the atomic
        mirror.  The action of the atomic mirror prevents the atom from reaching
        the surface and creates a microscopic trap.  The central contour graph
        shows the potential defining the quantum channel.  The two adjacent plots
        give the potential in a direction orthogonal to the charged wire and
        orthogonal / parallel to the mirror surface.  Distances are given from the
        location of the point charge and the surface of the atom mirror.}
        \protect{\label{f:qdot_potential}}
\end{figure}

As an example we will discuss here the simplest case of a point
charge on the surface of an atom mirror.  The point charge creates
an attractive $1/r^4$ interaction potential:
\begin{equation}
        V_{pol}(r) =-\frac{1}{8 \pi \epsilon_0} \alpha \frac{1}{r^4}
        \label{e:vpol_dot}
\end{equation}
together with the atomic mirror it will create a potential well
\beq \lb{e:vtrap_dot}
        V_{trap}(\vec{x})=U_m(x)+V_{pol}(r)
\eeq
that forms a microscopic cell for the atoms, such as that shown in
Fig.~\ref{f:qdot_potential}.  It can be viewed as the atom optical
analog to a quantum dot.

Far above the atom mirror the trapping potential $V_{trap}(\vec{x})$
(Eq.~\ref{e:vtrap_dot}) is dominated by $V_{pol}(r)$ (Eq.~\ref{e:vpol_dot})
and behaves like $1/r^4$, consequently there will be a {\em finite} number
of bound states.  For a specific set of parameters (weak potentials) there
will be {\em one} single bound state.

Typical parameters for such neutral atom quantum dots are given in
table~\ref{t:QuantDot}.  The potentials are not as deep as for the
quantum wires, but in general the discussion about the quantum wire
potentials given above also hold here.  Again using highly charged
dots the lower-lying eigenstates in the neutral atom quantum dots
can be localized to better than the wavelength of light and the
atoms will be trapped in the Lamb-Dicke regime.  With stronger atom
mirrors it is even conceivable to separate the bound states by more
than the linewidth of the optical transition allowing sideband
cooling similar to ions in ion traps \cite{iontraps}.

\section{Loading Atoms in Mesoscopic Surface Traps}
\lb{s:Loading}

An important question now arises, how to load atoms in these
microscopic traps or guides.  Atoms have first to be trapped and
cooled without coming in contact with the mirror surface.  The
evanescent wave mirror can repel all ground state atoms, and atom
traps close to or even at an evanescent wave mirrors were recently
demonstrated \cite{r:MOST,r:GOST}. Therefore loading atoms into
mesoscopic traps will be easier in the case of the evanescent wave
mirror based devices.  This is harder for the magnetic mirror,
since scattering light from atoms changes their internal magnetic
states, and the atoms in wrong magnetic substates will not be
reflected, but attracted to the mirror surface.  So loading schemes
like the one proposed by E.  Hinds \ea \cite{Hinds98} will have to
be used to transfer atoms from a MOT to a surface trap.

One loading scheme can be based on a modification of the magneto
optic surface trap (MOST) as demonstrated by J.  Mlynek's group in
Konstanz \cite{r:MOST}. A sizable number of atoms loaded and cooled
in a standard MOT at some distance to the surface and then shifted
fast (in a few ms) where they are then held in the region of the
mesoscopic traps or guides by the magneto optic surface trap. By
such a procedure, loading fare from the surface and then shifting
the atoms to te surface, one can transfer the density of a regular
MOT on to the surface. By simultaneously switching off the MOST and
switching on the charged structures one can trap some atoms in the
mesoscopic traps or guides.  The loading probability can then be
estimated by the product of trap (guide) volume times atom density.
Using a standard MOT density of $10^{11}\; cm^{-3}$ and a guide
cross section of $1 \; \mu m^2$ one can expect 100 trapped atoms
for 1 mm guide length.  One can think of enhancing the loading
probability by switching the electric trapping field on during the
MOST.  This will provide an additional potential minimum to attract
the cold atoms, similarly to loading atoms in a FORT \cite{FORT}.

A second promising loading scheme can be based on the gravito optic
surface trap \cite{r:GOST}.  There Atoms are loaded and cooled in a
small ($\mu m$ size) layer close to the surface of an evanescent
wave mirror. This is exactly the region where the atoms will then
be trapped in the mesoscopic traps. Switching on the trapping
electric fields during the final stage of the loading should again
enhance the probability to load the atoms in the small mesoscopic
traps.

One can also think of a two step process: First loading atoms in
one of our recently demonstrated atom guides created by a current
carrying wire \cite{WireGuide}.  These guides can transport the
atoms to the surface mounted atom optical devices.  Such a schema
has the advantage that it seems straight forward to load a BEC into
the wire guide which would then transport the atoms and connect
them modematched to the quantum wires.

\section{Outlook}
\lb{s:Outlook}

We have shown that by mounting changeable nanofabricated structures
on the surface of an atom mirror, one can design and create very
versatile microscopic and mesoscopic potential structures for
neutral atoms.  These neutral atom quantum wires and quantum dots
have many advantages compared to the other guides
\cite{Matter_Guides}:

\begin{itemize}
        \item {\em a)} The potential can be designed at will, using well-known
        technology.  It is easy to achieve very small guides, and single mode
        propagation or storage in the ground state should be possible.  In addition
        these guiding and trapping potentials can be easily modified by applying
        fields to additional electrodes.  Loading of the guides and
        traps seems feasible using known cooling and trapping technology.

        \item {\em b)} In contrast to the guiding and trapping in a hollow optical
        fiber \cite{AtomFiber} the neutral atom quantum wires and dots are in the open,
        above a surface, and the desired UHV environment, which is essential for
        coherent trapping or guiding can be easily realized.

        \item  {\em c)} Using use the well-developed nanofabrication techniques
        the principles of guiding can be easily extended to more
        complicated structures like a beamsplitter, an interferometer, or even
        integrated atom optical devices, like interferometers and inertial
        sensors or complex quantum networks for guided neutral atoms.  Furthermore
        additional electrodes located close to the wire can be used to modified the
        guiding potential on demand. One can easily imagine designing
        switches, gates, modulators etc. for guided atoms.

        \item {\em d)} Having the atoms trapped in microscopic traps near a surface
        will allow the integration of atom optics and light optics.  Waveguides
        for light can be nanofabricated on the atom mirror surface.  They can be
        used to address individual neutral atoms in quantum dots, and additional
        electrodes can be used to shift the atoms in and out of resonance.  This
        will allow the construction of integrated optics devices for atom light
        manipulation which might be used to build quantum registers for quantum
        communication and quantum computation \cite{r:NeutAtomQuComp}.

        \item {\em e)} Atom optics is inherently non linear allowing easy
        coupling between atoms in different channels of a waveguide network,
        just by using waveguide beam splitters.  If one can control the motion
        of the atoms in these guides to such an extend that one can send
        independent atoms simultaneously on a beam splitter,
        then one will be able to entangle independent atoms by their mutual interaction.
        This might allow to construct quantum computation gates and networks using atoms propagating in these
        waveguides \cite{r:NeutAtomQuComp}.

\end{itemize}

This work was supported by the Austrian Science Foundation (FWF),
project S06505, the Jubli\"{a}ums Fond der \"{O}sterreichischen
Nationalbank, project 6400, and by the European Union, contract Nr.
TMRX-CT96-0002.

\onecolumn

\newpage

\begin{table}
        \caption{Typical parameters for Li and Rb atoms guided in neutral atom
        quantum wires located above a typical evanescent wave mirror and a typical magnetic
        mirror.  Ground state properties are calculated using a harmonic oscillator
        approximation of the trapping potential near the minimum.  The
        $z$-direction is perpendicular to the mirror surface $x$-direction is
        parallel to the mirror surface and orthogonal to the wire.}
        \begin{center}
        \begin{tabular}{|c|c|cc|cc|cc|c|}
        \hline
        atom &wire&\mco{2}{c|}{potential}&\mco{5}{c|}{ground state properties}\\
                \cline{3-9}
            &charge& depth  &  distance
            &\mco{2}{c|}{frequency [kHz]}&\mco{2}{c|}{size [$\mu$ m]}&scat. rate\\
            &[pC]& [neV]  & [$\mu m$] & z & x & z & x &[kHz]\\
                \hline
        \mco{3}{|r}{evanescent wave mirror :}&
        \mco{2}{l}{$U_m$=1 $\mu$eV}&
        \mco{2}{l}{$\kappa = 0.1 \mu$m}&
        \mco{2}{l|}{$\Delta = 1000 \Gamma$}\\
        \hline
        Li & 0.33 &-5.05 & 0.60 & 211 & 125 & 0.083  & 0.107 & 4.6      \\
        Rb & 0.22 &-4.47 & 0.610 & 57 & 33 & 0.046  & 0.060 & 5.2       \\
        \hline
        \mco{3}{|r}{magnetic mirror :}&
        \mco{3}{l}{$U_m$=6.4 $\mu$eV  (B$_0$=1100 G)}&
        \mco{3}{l|}{$\kappa = 1.5 \mu$m  (9.5 $\mu$m magnetisation)}\\
        \hline
        Li & 1.05 &-0.063 & 20 & 1.15 & 0.37 & 1.12  & 1.98 & - \\
        Li & 15.7 &-75.5 & 7.2 & 56 & 43 & 0.161  & 0.183 & - \\
        Rb & 0.43 &-0.017 & 22 & .166 & .05 & 0.85  & 1.55 & - \\
        Rb & 10.8 &-64.9 & 7.5 & 14.8 & 10.8 & 0.090  & 0.105 & - \\
        \hline
        \end{tabular}
        \end{center}
        \protect\label{t:QuantWire}
\end{table}

\begin{table}
        \caption{Typical parameters for Li and Rb atoms trapped in neutral atom
        quantum dots above a typical evanescent wave mirror and a typical magnetic
        mirror.  Ground state properties are calculated using a harmonic oscillator
        approximation of the trapping potential near the minimum.  The
        $z$-direction is perpendicular to the mirror surface $x$-direction is
        parallel to the mirror surface and orthogonal to the wire.}
        \begin{center}
        \begin{tabular}{|c|c|cc|cc|cc|c|}
        \hline
        atom&point&\mco{2}{c|}{potential}&\mco{5}{c|}{ground state properties}\\
                \cline{3-9}
            &charge&depth&distance&\mco{2}{c|}{frequency [kHz]}
            &\mco{2}{c|}{size [$\mu$ m]}&scat.rate\\
            &electrons& [neV]  & [$\mu m$] & z & x & z & x &[kHz]\\
                \hline
        \mco{3}{|r}{evanescent wave mirror :}&
        \mco{2}{l}{$U_m$=1 $\mu$eV}&
        \mco{2}{l}{$\kappa = 0.1 \mu$m}&
        \mco{2}{l|}{$\Delta = 1000 \Gamma$}\\
        \hline
        Li & 141 &-0.59 & 0.72 & 88 & 60 & 0.13  & 0.16 & 1.5   \\
        Rb & 112 &-0.99 & 0.65 & 32 & 27 & 0.061  & 0.067 & 2.6 \\
        \hline
        \mco{3}{|r}{magnetic mirror :}&
        \mco{3}{l}{$U_m$=6.4 $\mu$eV  (B$_0$=1100 G)}&
        \mco{3}{l|}{$\kappa = 1.5 \mu$m  (9.5 $\mu$m magnetisation)}\\
        \hline
        Li & 10000 &-0.005 & 23 & 0.42 & 0.13 & 1.86  & 3.3 & - \\
        Li & 80000 &-3.2 & 11 & 14 & 9 & 0.33  & 0.41 & - \\
        Rb & 10000 &-0.012 & 21 & .19 & .07 & 0.78  & 1.34 & - \\
        Rb & 63000 &-5.0 & 10 & 4.9 & 3.8 & 0.16  & 0.18 & - \\
        \hline
        \end{tabular}
        \end{center}
        \protect\label{t:QuantDot}
\end{table}


\begin{thebibliography}{99}
\bibitem{EvWaveMirror}
        For an evanescent wave mirror \cite{Cook,Balykin,Esslinger,Seifert,EvMirrProblem}
        the length scale $\kappa$ is given by:
        $\kappa = 2 k_L \sqrt{n^2 \sin^2 \theta_i - 1}$ where $k_L$ is the wave
        vector of the light, $n$ is the refractive index of the glass and
        $\theta_i$ the incident angle of the laser beam.  To obtain coherent
        reflection the light has to be far-detuned and the surface has to be
        of very high quality.  In general Eq.~\ref{e:at_mirror} will be only an
        approximation due to the imperfections in the mirror surface and the
        resulting interference between the evanescent wave and light scattered
        at these imperfections.
\end{thebibliography}
\end{document}